\documentclass[12pt]{article}
\begin{document}
\title{ {\bf  Correlation functions and dissipation \\ in hot QCD} }
\author{F. Guerin \\ 
  Institut Non Lineaire de Nice,  1361 route des Lucioles, 06560 Valbonne, France } \date{}
\maketitle
\begin{abstract}
A recently proposed generating functional allows the construction of the full set of n-point Green functions in QCD at high temperature and at distances larger than 1/gT. One may then learn how the system maintains its thermal equilibrium in the quantum field theory approach,  i.e. which process compensates for the important dissipation due to collisions. This system may be characterized by quantities which have a classical limit. One finds that the fluctuations of the colour field are not gaussian ones. A comparison is made with B\"{o}deker's classical approach where a random noise is the source of fluctuations.   
\end{abstract}
 
\section{Introduction}
QCD at high temperature possesses several energy scales: $T, gT, g^2T\ln 1/g$  \ldots , where $g(T)\ll 1$. Two effective theories have been constructed, each one associated with a specific scale. In the HTL effective theory \cite{Pi1,FTay}, the scale $T^{-1}$ is integrated out. Its features and consequences are well established \cite{MLB,BI4}. In the ultrasoft effective theory, the scales $T^{-1}$ and $(gT)^{-1}$ are integrated out \cite{Bod1}. Those effective theories exhibit semi-classical features and may be obtained via different approaches (quantum theory, response function, transport equations, distribution functions) \cite{Bod2,BI1,AY2,Lit}. At the scale $(g^2T)^{-1}$, the physics is expected to be non-perturbative, as thermal perturbation theory breaks down because of strong nonlinear effects.

As one goes to larger (space) scales, the importance of dissipation at this scale  increases, as energy is transferred towards smaller scales. In the HTL case, the damping only occurs in some kinematical regions (space-like gluons). The quantum theory is mostly constructed out of those regions, which may then be reached via analytical continuations in the energy variables. In the ultrasoft case, the damping is of the same order as the momentum, and it occurs in the whole energy domain.  The approaches split into two branches, the classical approach \cite{Bod1,Bod2,AY2,Lit} and the quantum approach \cite{BI1,BI4}, which differ in the resulting set of classical equations for the effective field. 

The original derivation, by D. B\"{o}deker, is within the framework of the classical HTL effective theory. The fields in this theory are split into soft $p\sim  gT$ and ultrasoft $p\sim g^2T$, and the soft fields are integrated out to obtain the following set of effective equations for the ultrasoft field
\begin{equation}
[D_{\mu} , F^{\mu \nu}] (X) = m_D^2 \int_v \ v^{\nu} W(X,{\mathbf v})  \label{aa1}
\end{equation}
\begin{equation}
(v.D_x+\hat{C}) W(X,{\mathbf v}) = {\mathbf v}.{\mathbf E}(X) + \xi(X,{\mathbf v})    \label{aa2}
\end{equation}
where $v.D_x$ is a gauge-covariant drift operator with $v^{\nu}(v_0=1,{\mathbf v}), ({\mathbf v}^2=1)$ , $\int_v=\int {d \Omega_v /{4\pi}}$, and  $\hat{C}$ is a collision operator local in $X$, but non-local in $\mathbf v$. \  $\xi(X,{\mathbf v})$ is a gaussian white noise term, which compensates for the dissipation arising from the collision term, so that the system is in thermal equilibrium. The gluons $p\sim T$ take part in the collective motion $W(X,{\mathbf v})$, whereas the gluons $p\sim gT$ are exchanged during the collisions that $\hat{C}$ represents.

Alternatively, in the quantum approach, one considers the response of the system to a weak external perturbation, i.e. to a source $j_{\mu}^{ext}(X)$ which induces large-scale inhomogeneities. The response is an induced current $j_{\mu}^{ind}(X)$, and the resulting set of classical equations is
\begin{equation}
[D_{\mu} , F^{\mu \nu}] (X) =  j_{\mu}^{ext}(X) + j_{\mu}^{ind}(X) \label{aa3}
\end{equation}
\begin{equation}
j_{\mu}^{ind}(X)= m_D^2 \int_v \ v^{\nu} W'(X,{\mathbf v})  
\end{equation}
\begin{equation}
(v.D_x+\hat{C}) W'(X,{\mathbf v}) = {\mathbf v}.{\mathbf E}(X)    \label{aa4}
\end{equation}
where $\hat{C}$ is the same as in Eq.(\ref{aa2}). These equations are obtained from the Schwinger-Dyson equation for the two-point function via a gradient expansion. In this approach, the response to a short perturbation dies out because of the damping due to $\hat{C}$. In order to draw conclusions about the system in thermal equilibrium, one has to refer to the fluctuation-dissipation theorem, which is generally obeyed by the two-point function.

Morever, this induced current is the generating functional of the $n$-point response functions. Recently, a generating functional has been presented \cite{FG4} that allows one  to obtain the full set of $n$-point amplitudes, i.e. it contains all the information on the (effective) quantum system at a scale larger than $(gT)^{-1}$ and smaller than $(g^2T)^{-1}$ . As the system is in thermal equilibrium, there exists a mechanism which compensates for the energy dissipation occurring at this scale. In macroscopic physics, the compensation is attributed to thermal fluctuations. The purpose of this paper is  to characterize, tentatively, the process that maintains the system in equilibrium, and to compare it to the statistical physics  approach.\\

In quantum field theory at finite temperature $T$, a doubling of the degrees of freedom occurs,  compared  with field theory at $T=0$. There are several formalisms available \cite{MLB}. The Keldysh basis \cite{Chou,Kel}, a rotated version of the Closed-Time-Path formalism, offers two particular advantages for our purpose: (i) the generating functional of ultrasoft QCD is simple (ii) there exist $n$-point Green functions which are strong candidates for a direct connection with the correlation functions of statistical physics, at high $T$. \\
Section \ref{sec2} summarizes several general features of the (unfamiliar) Keldysh basis: (i) the relation between one-particle-irreductible (1PI) vertices and connected Green funtions, (ii) the relation between Keldysh Green functions and expectation values of ordered product of field operators $A_{\mu}(X)$.  A reader familiar with this basis may go directly to Sec.\ref{sec23},  where candidates for an identification with correlation functions are selected. Section \ref{sec3} is devoted to the quantum approach to ultrasoft QCD. At the scale $(g^2T \ln 1/g)^{-1}$  (with $\ln 1/g \gg 1$), the connected Green functions of interest are deduced from the 1PI vertices and these are obtained from the generating functional. One result is that the fluctuations of the field $A_{\mu}(X)$ at high $T$ are not gaussian. The HTL case is examined as a limiting case, and simple explicit expressions are obtained. In Sec.\ref{sec4}, those results are compared with those from  the classical approach, i.e. from Eqs.(\ref{aa1})(\ref{aa2}),  where a gaussian noise compensates for the dissipation.  Section \ref{sec5} contains a summary of the result and discusses an effective action for ultrasoft QCD.

\section{The n-point Green functions in Keldysh basis}
\label{sec2}
In the functional approach to field theory at temperature $T$, the time contour that enters the action has to run forwards and backwards between $t=-\infty$ and  $t=+\infty$. When expressed as a single time integral, the result is the doubling of the degrees of freedom, as the field $A_{+}$ lives on the forward branch and the field $A_{-}$ on the backward one. In consequence, to an $n$-point amplitude are associated $2^n$ different components. The familiar representation involves $A_{+}$ and $A_{-}$  and the corresponding Green functions are expectation values of mixed combinations of time-ordered and anti-time-ordered products of operator fields. In the Keldysh basis \cite{Chou}, the fields are $A_{+}\pm A_{-}$, and the Green functions are expectation values of combinations of time-ordered commutators and anticommutators  of the operator fields. Among them are the response functions, i.e. the retarded amplitudes. \\
The present state, in the quantum approach to ultrasoft QCD, follows. From the knowledge of the retarded $n$-point functions, consistency condition and gauge symmetry have lead to the whole set of 1PI vertices in the Keldysh basis. We now need to construct the connected Green functions from the 1PI vertices, and to identify the relevant ones for our purpose. 

\subsection{Relating connected amplitudes to 1PI vertices}
\label{sec21}

It will be useful to obtain the results in two ways: (i) the naive approach is to go from the Keldysh basis to the Retarded-Advanced ($R/A$) basis, where the relation is immediate, (ii) the other approach is to extract the prescription from the general Legendre transform in the Keldysh basis. 

 Our notation is $P_i (p_i^0, {\mathbf p}_i)$. In the Keldysh basis, the indices are 1 and 2, the 1PI  $n$-point vertices are $K_{i j \cdots n}(P_1, P_2 , \cdots P_n)$ and the connected amplitudes are $G^c_{i j \cdots n}(P_1, P_2 , \cdots P_n)$ . All momenta are incoming $\sum^n_{i=1} P_i =0$. In the $R/A$ basis, a leg may be of type $R$ or of type $A$, and the 1PI amplitudes are $\Gamma_{RA\cdots A}$. In this basis, the relation between  amputated and non-amputated functions is straightforward. Indeed, to a leg of type $R$ one attaches a retarded (dressed) propagator $G_r(P)$, and to a leg of type $A$  an advanced one $G_a(P)$.  \\
The connection between 1PI $n$-point vertices in the $R/A$ basis and in the Keldysh basis has been studied in Ref.\cite{vWK,FG4}. Here we just quote the results and make use of them. One defines 
\begin{equation}
N(P_i)=n(p_i^0)+{1\over 2} = {1\over 2}\coth {\beta\over 2}p_i^0
\label{a1}
\end{equation}
\begin{equation}
G_c(P) = N(P) [ \ G_r(P)-G_a(P)\ ]
\label{a2}
\end{equation}
where $n(p_i^0)$ is the Bose statistical weight. One relation between the bases is 
\begin{equation}
K_{21 \cdots 1}(P_1,P_2 \cdots P_n)=\Gamma_{RA \cdots A}(P_1,P_2 \cdots P_n) 
\label{a3}
 \end{equation}
so that
\begin{eqnarray}
G^c_{21 \cdots 1}(P_1,P_2 \cdots P_n)&=&G_r(P_1)G_a(P_2)\cdots G_a(P_n)\ \Gamma_{RA \cdots A}(P_1,P_2 \cdots P_n)  \nonumber \\
G^c_{21 \cdots 1}(P_1,P_2 \cdots P_n)&=&G_r(P_1)G_a(P_2)\cdots G_a(P_n)\  K_{21 \cdots 1}(P_1,P_2 \cdots P_n)  \label{a3b}
\end{eqnarray}
i.e. the relation is now written in the Keldysh basis. Similarly
 \begin{eqnarray}
K_{221 \cdots 1}(P_1,P_2 \cdots P_n)=\Gamma_{RRA \cdots A}(P_1,P_2 \cdots P_n)  \nonumber \\
- \ N(P_1) \ \Gamma_{ARA \cdots A} \  - N(P_2) \  \Gamma_{RAA \cdots A}
\label{a4} \end{eqnarray}
\begin{eqnarray}
G_{221 \cdots 1}^c=G_a(P_3) \cdots G_a(P_n) [\ G_r(P_1)G_r(P_2)\Gamma_{RRA \cdots A} \nonumber \\
-  N(P_1) G_a(P_1)G_r(P_2)\  \Gamma_{ARA \cdots A} \  - N(P_2) G_r(P_1)G_a(P_2)\  \Gamma_{RAA \cdots A}  \ ]
 \end{eqnarray}
\begin{eqnarray}
\lefteqn{G_{221 \cdots 1}^c=G_a(P_3) \cdots G_a(P_n) [\ G_r(P_1)G_r(P_2)K_{221 \cdots 1} \nonumber} \\ 
& &  + G_c(P_1)G_r(P_2)\  K_{121 \cdots 1} \  +  G_r(P_1)G_c(P_2)\  K_{211 \cdots 1}\  ]
\label{a5} \end{eqnarray}
One notices the occurrence of $G_c(P_i)$ , which will play an important role later on. Similarly, the next relation is
\begin{eqnarray}
G_{2221 \cdots 1}^c=G_a(P_4) \cdots G_a(P_n) \  \lbrace  \ G_r(P_1)G_r(P_2)G_r(P_3)K_{2221 \cdots 1} \nonumber \\ 
+ [ \ G_c(P_1)G_r(P_2)G_r(P_3)\  K_{1221 \cdots 1} \    +  ( 1\leftrightarrow  2)  + (1\leftrightarrow 3)\ ]
\nonumber \\
+  [ \ G_r(P_1)G_c(P_2)G_c(P_3)\  K_{211 \cdots 1}  +  ( 1\leftrightarrow  2)  + (1\leftrightarrow 3)\ ] \ \ \rbrace
\label{a6} \end{eqnarray} 
where $(1\leftrightarrow 3)$ means the term obtained in the exchange of all variables of legs 1 and 3.
The denominators of the propagators $G_r(P)$ and $G_a(P)$ are respectively $(p_0+i\epsilon)^2 - {\mathbf p}^2 -\Pi_r(P)$ and $(p_0-i\epsilon)^2 - {\mathbf p}^2 -\Pi_a(P)$, with $\Pi_a(P)=\Pi_r^*(P)$. \\

Alternatively, one may obtain those relations via the Legendre transform in the Keldysh basis (see Sec.\ref{sec22}). The result is, for $n\geq 3$
\begin{equation}
G_{ijk\cdots n}^c = M_{i i'}(P_1)M_{j j'}(P_2) \cdots M_{n n'}(P_n) \ K_{i'j'k'\cdots n'}
\end{equation}
where the 2x2 matrix $M(P)$ is
\begin{equation}
 M(P) = \begin{array}{|cc|}
G_a(P) &  0 \\ G_c(P) & G_r(P) 
\end{array}    
\end{equation}  
The zero of the matrix M ensures that $K_{11\cdots 1}=0$ leads to $G_{11\cdots 1}=0$. Both are important features of the Keldysh basis. Also, the retarded amplitudes $G_{211\cdots 1}^c$ are simply related to $K_{211\cdots 1}$. The $2$-point function is written as a 2x2 matrix
\begin{equation}
\sigma_1 G(P) \sigma_1 H(P) =1  \ \ \ , \ \ \ M(P)= G(P) \sigma_1
\label{a8} \end{equation}
\begin{eqnarray}
H_{21}=(p_0+i\epsilon)^2 -p^2-\Pi_r(P) \ \ \ & , &\ \ \ H_{12}=(p_0-i\epsilon)^2 -p^2-\Pi_a(P)  \nonumber
\\ H_{11}=0 \ \ \   & , &  \ \ \ H_{22} = N(P)[ H_{21} -H_{12} ]
\label{a8b}
\end{eqnarray}
$\sigma_1$ is the Pauli matrix,  and it arises from the fact that, in the Keldysh basis, the source term in the Legendre transform reads (see Eq.(\ref{a10}))
\begin{equation}
\int \  J_1 A_2 + J_2 A_1 = \int \  ( A_1 \  A_2) \  \sigma_1 \left( \begin{array}{c} J_1 \\ J_2  \end{array}\right)
\end{equation}
i.e. the generating functional $W$ is a function of $\sigma_1 J$. From Eq.(\ref{a8}) one deduces
\begin{equation}
G_{21}=G_r=H_{21}^{-1} \ \ , \ \ G_{12}=G_a=H_{12}^{-1} \ \ , \ \  G_{22}=G_c=-G_rH_{22}G_a
\end{equation}
so that Eq.(\ref{a8b}) translates into $G_c(P) = N(P) (G_r-G_a)$. 

For $n\geq 4$, it is well known that the connected Green functions contains other terms in addition to those written in Eqs.(\ref{a5})(\ref{a6}). They arise from one-particle-reducible pieces, i.e. from 1PI vertices linked with dressed propagators. For example, for the 4-point case, two 3-point vertices are linked with one propagator, and the prescription is
\begin{eqnarray}
G_{ijkl}^{c(2)} = M_{i i'}(P_1)M_{j j'}(P_2) K_{i'j'm} (H^{-1})_{m m'}(P_1+P_2) K_{m'k'l'} M_{k'k}(P_3)M_{l'l}(P_4)  \nonumber \\
 + (2\leftrightarrow  3) + (2\leftrightarrow  4)   \label{a9}
\end{eqnarray}

\subsection{The Green functions in position space}
\label{sec22}
As the Keldysh basis is unfamiliar, since rarely used, we summarize the properties that are needed to fully understand the physical features of our results on ultrasoft QCD. Those properties are taken from Ref.\cite{Chou} (our definition of $G_c(P)$ differs by a factor of 2, see \cite{vWK}).\\
The same generating functional can generate $n$-point functions in different representations of the Closed-Time-Path formalism, provided that the external source term is expressed in the corresponding functional arguments. This source term writes
\begin{equation}
\int_X [ \ j_+(X) A_+(X)-j_-(X) A_-(X) \ ] = \int_X \ [ J_2A_1+J_1A_2 ]
\end{equation}
with $\int_X=\int_{-\infty}^{+\infty} d t \int d_3x$, where $A_+$ is the component of the $A$ field along the forward branch of the time contour, $A_-$ the component along the backward branch, and
\begin{equation}
A_1 ={(A_++A_-)/2} \ \ , \ \ A_2=A_+-A_- \ \ , J_1={(j_++j_-)/ 2} \ \ , \ \ J_2= j_+-j_-
\end{equation}
In the Keldysh basis, the Legendre transform reads
\begin{equation}
W(J_1,J_2)= \int_X   [J_1A_2+J_2A_1] \ + \ \Gamma(A_1, A_2)
\label{a10}
\end{equation}
\begin{equation} 
{\partial \Gamma\over{\partial A_1(X)}} = -J_2(X) \ \ , \ \ {\partial \Gamma\over{\partial A_2}} = -J_1  \ \ , \ \ {\partial W\over{\partial  J_1}} = A_2 \ \ , \ \ {\partial W\over{\partial J_2}} = A_1
\end{equation}
\begin{equation}
\Gamma(A_1,A_2)=\sum_{n=2}^{\infty} {1\over n!} \int_{all \ X}\ K_{ijl\cdots n}(X_1, \cdots X_n)  \ A_i(X_1)A_j(X_2) \cdots A_n(X_n)
\end{equation}
\begin{equation}
G_{ij\cdots n}^c(X_1 \cdots X_n) = {\partial W \over{ \partial J_i\partial J_j \cdots \partial J_n}} \ |_{J_1=J_2=\cdots=0}
\end{equation}
Since $K_{11\cdots1}=0$, $\Gamma(A_1,A_2)$ (and all its derivative with respect to $A_1$ only) vanishes for $A_2=0$. Setting $J_2=0$ amounts to $A_2=0$. Two consequences are important features of the Keldysh basis: (i) $W(J_1, J_2=0)=0$ ,  (ii) $A_2=0$ is $A_+(X)=A_-(X)$, i.e. after differentiation of $W$, setting $J_2=0$ says that the field $A$ takes the same value on both branches of the complex time path. \\

The comparison of the Green functions in the bases $(A_+,A_-)$ and  $(A_1,A_2)$,  lead to the following results \cite{Chou}. The amplitudes $G_{211\cdots1}$ are the usual retarded Green functions, which involve commutators of the field $A_{\mu}(X)$. A functional derivative with respect to $J_1$ generates a commutator, a derivative with respect to $J_2$ generates an anticommutator. For example
\begin{equation}
G_{21}=-i \theta(1,2) \langle  \ [A(1),A(2)]  \ \rangle  \ \ \ , \ \ \ 
 G_{22}=-{i\over 2} \langle \  \lbrace A(1),A(2) \rbrace \  \rangle  \label{a12}
\end{equation}
\begin{eqnarray}
\lefteqn{
 \ \ \ \ \  G_{211}=(-i)^2 \theta(1,2,3) \langle \  [ \ [A(1),A(2)], A(3) ] \  \rangle + (2 \leftrightarrow 3)  \nonumber} \\ &  &
G_{221}=(-i)^2 {1\over 2} \left( \theta(1,2,3) \langle \  [ \lbrace A(1),A(2)\rbrace , A(3) ]  \ \rangle +  \theta(1,3,2) \langle \ [  \lbrace A(1),A(3)\rbrace, A(2) ]  \ \rangle \right) + (1\leftrightarrow 2)   \nonumber \\  & & 
G_{222}=(-i)^2 {1\over 4}  \theta(1,2,3) \langle  \ \lbrace  \lbrace A(1),A(2)\rbrace, A(3)  \rbrace  \ \rangle + 5  \ {\mathrm{perm}}   \label{a15}
\end{eqnarray}
where $\theta(i,j,k)=\theta(i,j)\theta(j,k)$, and one sums over all permutations of $1,2,3$ in $G_{222}$ . More generally
\begin{equation}
G_{22\cdots 2}=\sum_{perm}{(-i)^{n-1} \over 2^{n-1}}  \theta(1,2 \cdots n)) \langle  \ \lbrace  \lbrace A(1),A(2)\rbrace, \cdots A(n) \rbrace  \ \rangle 
\label{a16}
\end{equation} 

\subsection{The high temperature limit}  \label{sec23}
One considers our case of interest, i.e. an effective quantum field theory for ultrasoft (or soft) gluons where $p_0\ll T$, then the Bose factor 
\begin{equation}
N(P) =n(p_0) + {1\over 2} = {1\over 2} \coth {\beta\over 2}p_0 \approx {T\over p_0}  \gg 1
\end{equation}
The occupation number ${\cal N}$ is so large than ${\cal N}+1 \approx {\cal N}$, and one expects the commutators of the fields to be negligible in front of anticommutators, in the classical limit \cite{Chou}. One concludes that the amplitudes $G_{22\cdots2}$ of Eq.(\ref{a16}) should correspond to the correlation functions of the statistical classical theory, i.e. the connected amplitudes $G_{22\cdots2}^c$ should correspond to the cumulants of the correlation functions and they should characterize the fluctuations. The overwhelming role of the amplitudes with only indices $2$ is explicit in the comparison of the factors that multiply $K_{21\cdots1}$ in Eqs.(\ref{a3b})(\ref{a5})(\ref{a6}). Indeed, each extra index $2$ brings an extrafactor $T/p_0$, since this index appears with a factor $G_c(P)\approx (G_r(P)-G_a(P))T/p_0$.

In  Appendix \ref{App1}, we discuss the generalization of the fluctuation-dissipation theorem  to the $n$-point functions' case.

\section{Ultrasoft QCD}
\label{sec3}

 As summarized in the introduction, ultrasoft QCD is an effective theory for gluon momenta $p\ll gT \ll T$.  The gluons $p\sim T$ take part in a collective motion, whereas the gluons $p\sim gT$ are exchanged during collisions which change the direction $\mathbf v$ of the collective motion.  This section is devoted to the quantum approach which uses the response of the system to an external perturbation (see Eqs.(\ref{aa3})(\ref{aa4}) and comments that follow). At a space scale larger than $(gT)^{-1}$, the dissipation is important, and our aim is to characterize the process that maintains the system in thermal equilibrium in this approach.  We will  examine the 1PI vertices in the Keldysh basis and the connected Green functions $G_{22..2}^c$ of this quantum theory, either at the scale $p\sim g^2T \ln 1/g$ (i.e. an intermediate scale between $gT$ and $g^2T$), or at the scale $p\sim gT$.\\
 The equilibrium condition sets constraints on the $n$-point gluon amplitudes. The ultrasoft effective theory satisfies them in an economical way in the Keldysh basis: (i) all 1PI amplitudes with more than two indices 2 vanish, (ii) there is a generating functional $j_{\mu}(X, A)$ for the amplitudes with one index 2 \cite{BI1}, and another functional ${\cal K}_{\mu \nu}(X, Y, A)$ for the amplitudes with two indices 2 \cite{FG4}. $j_{\mu}=j_{\mu}^{ind}$ is the response of the system to an external perturbation, and ${\cal{K}}_{\mu \nu}$ would vanish if there were no dissipation. They are both expressed in terms of one Green function
\begin{equation}
i\ (v.D_x + \hat{C}) \ G_{ret}( X, Y; A ;{\mathbf v}, {\mathbf v'})  = 
 \delta^{(4)}(X-Y) \ \delta_{S_2} ({\mathbf v} - {\mathbf v'})
\label{b1} 
\end{equation}
\begin{equation}
\hat{C} \  G_{ret}( X, Y; A ;{\mathbf v}, {\mathbf v'})  =\int_{v"}  \ C({\mathbf{v}, \mathbf{v"}}) 
G_{ret}( X, Y;A ;{\mathbf v"}, {\mathbf v'})
\label{b2}
\end{equation}
with $ G_{ret} ( X, Y)  =  0$  for $ X_0 < Y_0$ , $\int_v=\int {d \Omega_v /{4\pi}}$ , $v^{\mu}(v_0=1,{\mathbf v}), ({\mathbf v}^2=1)$ , and  $D_x$ is the covariant derivative  $D_{\mu}=\partial_{\mu} + i g T^a A^a_{\mu}(X)$ with $T^a$ in the adjoint representation. The collision operator $\hat{C}$  is symmetric in ${\mathbf v}$ and ${\mathbf v'}$ and has positive eigenvalues of order $g^2T \ln 1/g$, except for a zero mode. Its precise form is irrelevant for our purposes. To leading log accuracy, the form of $\hat{C}$ has been derived in the different approaches \cite{Bod1,Bod2,AY2,BI1,BI4,Lit}.   Attempts to go beyond this may be found in Ref.\cite{AY1,FG3}. 

Eq.(\ref{b1}) for $G_{ret}$ may be solved by recurrence as an expansion in powers of $C({\mathbf{v}, \mathbf{v'}})$. It leads to the following interpretation. A collision changes the direction of $\mathbf{v}$. Between two collisions, $G_{ret}( X, Y; A ;{\mathbf v}, {\mathbf v'})$ describes a straightline propagation (in direction $\mathbf{v}$) in the background non-abelian field $A_{\mu}$ \cite{BI3}. Alternatively, the solution to Eq.(\ref{b1}) may be written as an expansion in powers of $A_{\mu}$, i.e. between two interactions with the $A$ field  there may occur many collisions which change the direction of ${\mathbf v}$.
 \begin{eqnarray} 
\lefteqn{ j_{\mu}^{a}(X; A) =  m_D^2[- g_{0\mu} A_0^a(X)  \nonumber} \\ & &
 + i \  \int_{v,v'}  \int d^4Y \ v_{\mu} \ 
  G_{ret}^{ab}( X, Y ; A ; {\mathbf v}, {\mathbf v'})\  \partial_{Y_0} (v'.A^b(Y)) \ ]
\label{b3}
  \end{eqnarray}
\begin{eqnarray}
\lefteqn{{\mathcal K}_{\mu \nu}^{a b}(X,Y; A) =m_D^2 \ T  \nonumber }\\  & &
\int_{v,v'}[\  v_{\mu}G_{ret}^{ab}(X,Y; A;{\mathbf v}, {\mathbf v'}) v_{\nu}' \ + \ v_{\nu}G_{ret}^{ba}(Y,X; A;{\mathbf v}, {\mathbf v'}) v_{\mu}' \ ]   \label{b3b}
\end{eqnarray}
where $m_D^2=g^2NT^2/3$. Functional derivation gives the $n$-point 1PI vertices
\begin{equation}
 g^{n-2} K_{2 1 \cdots 1}(X ; X_1 \cdots X_{n-1}) = 
 {\partial^{n-1} \over{\partial A(X_{n-1}) \cdots
 \partial A(X_1)}} \ \ j_{\mu}(X; A) \ \vert_{A=0}   \label{b3c}
 \end{equation}
$K_{2 2}(X,Y)={\cal{K}}_{\mu \nu}(X,Y; A=0)$, $K_{221}$ is obtained from one functional derivative of 
${\cal{K}}_{\mu \nu}$ and so forth. 
An important identity  is \cite{BI3,FG3}
 \begin{eqnarray}
 \lefteqn{{\partial G_{ret}( X, Y ; A ; {\mathbf v}, {\mathbf v'}) \over{\partial A_c^{\rho}(X_1)}}  \nonumber} \\  & &  =
 \int_{v"} G_{ret}( X, X_1; A ;{\mathbf v}, {\mathbf v"}) \ g T^c
 {v"}_{\rho} \ G_{ret}( X_1, Y; A ;{\mathbf v"}, {\mathbf v'})  \label{b3d}
 \end{eqnarray}
The bare vertices should be added to those effective vertices. However, they are subleading for momenta $p\ll gT$, as  it will be shown explicitely in Sec.\ref{sec4}. In the following, we concentrate on these 1PI vertices. Such an expansion of $G_{ret}( X, Y ; A ; {\mathbf v}, {\mathbf v'})$ in powers of $A$ is only meaningful for distances such that $(v.D + \hat{C})^{-1}$ can be expanded in powers of $A$. We need to assume $\ln 1/g \gg 1$ so that the scales $g^2T\ln1/g$ and $g^2T$ are well separated. For an ultrasoft field $gA\sim g^2T$, and we shall sit at the scale $\partial_x\sim g^2T\ln 1/g \gg gA$. As an aside comment, these 1PI vertices may be used to take into account the contribution of loops with momenta $g^2T\ln 1/g$ to the amplitudes at the scale $g^2T$. The resulting term in the collision operator $\hat{C}$ is local in $X$ only for $\ln 1/g \gg 1$.

\subsection{The ultrasoft case}
In this section, expressions for the connected Green functions are derived for momenta $p\sim g^2T \ln 1/g$, mostly concentrating on those with only indices 2.

For the 2-point functions, the functional derivatives lead to the following results
\begin{equation}
K_{21}(P, -P) = K_{21}(P) = m_D^2 \delta^{ab}\ [\int_{v,v'} v_{\mu} (v.P + i \hat{C})^{-1}  {v}_{\nu}'  \ (p_0) -  g_{\mu 0} g_{\nu 0} \  ]
\label{b5}
\end{equation}
$K_{12}$ is obtained  from $K_{21}$ with the following substitution $\mu \leftrightarrow \nu,  P \leftrightarrow -P$. 
\begin{equation}
K_{22}(P)  = m_D^2  \ T\  \delta^{ab}  \int_{v,v'} \ v_{\mu} (v.P + i \hat{C})^{-1}  {v}_{\nu}'    + v_{\nu} (- v.P + i \hat{C})^{-1}  {v}_{\mu}' 
\label{b6}
\end{equation}
so that  $K_{22}=(K_{21}-K_{12})T/p_0$, which is relation (\ref{a8b}) in the approximation $N(p_0)\approx T/p_0$. $K_{22}$ exists in the whole kinematical domain $(p_0,p)$, with the constraint $p_0, p \ll gT$. Ward identities relate the components \cite{FG4}  $P_{\mu}K_{21}^{\mu \nu}=0=P_{\mu}K_{22}^{\mu \nu}$. An explicit expression for $K_{21}$ exists in terms of continued fractions \cite{AY1,FG3}. 

The definition of the retarded and advanced propagators $G_r(P)$ and $G_a(P)$ requires the choice of a gauge and the projectors ${\cal{P}}^{\mu\nu}(P)$ on transverse and longitudinal components. They are the standard ones and they will not be needed. The denominators are respectively $P^2-\Pi_r(P)$ and $P^2-\Pi_a(P)$ with $K_{21}(P)=\Pi_r(P), \ K_{12}(P)=\Pi_a(P)$. The following useful relation is gauge-independent
\begin{eqnarray}
G_{22} = G_c^{\mu \nu} = {T\over p_0} \ [ \ G_r^{\mu\nu}(P) - G_a^{\mu\nu}(P)\ ] \nonumber \\ 
=  {T\over p_0}  \ G_r^{\mu\rho}(P)[ \Pi_{r \, \rho\sigma}(P)- \Pi_{a \, \rho\sigma}(P) \ ]G_a^{\sigma\nu}(P)
\label{b6b}
\end{eqnarray}
\begin{equation}
\Pi_a(P)=\Pi_r^*(P)=\Pi_r(-P)
\end{equation} \\

Turning to the 3-point amplitudes, all of them may be expressed in terms of the same quantities ${\cal{T}}(i,j,k)$.  If $(P_1, \mu, a)$ refers to the momentum, polarisation, colour of gluon 1,  $(P_2, \nu, b)$ refers to gluon 2, $(P_3, \rho, c)$ to gluon 3, one defines the ``time-ordered'' quantity
\begin{equation}
{\cal{T}}(1,3,2)=m_D^2 \int_{v,v',v"} v_{\mu}(v.P_1+i\hat{C})^{-1} v_{\rho}' \ (-v.P_2+i\hat{C})^{-1} v_{\nu}"
\label{b7}
\end{equation} 
\begin{equation}
{\cal{T}}(1,3,2)^* = {\cal{T}}(2,3,1) \label{b7b}
\end{equation}
$P_1+P_2+P_3=0$. The order in $v$ space and in colour space ($i f^{acb}$) follow the ``time-order''. The 1PI amplitudes have either one index 2 (they are the retarded amplitudes), or two indices 2, as $K_{222}=0$. For example,
\begin{equation}
K_{211}(P_1,P_2,P_3)= i f^{abc} {\cal{T}}(1,2,3)(-p_3^0) +  i f^{acb} {\cal{T}}(1,3,2)(-p_2^0)
\label{b8b}
\end{equation} 
\begin{equation}
K_{221}(P_1,P_2,P_3)= i f^{acb} T  \ [\  {\cal{T}}(1,3,2)- {\cal{T}}(2,3,1) \ ]
\label{b8}
\end{equation}
The  minus sign in $K_{221}$ comes from the colour factor $f^{bca}=-f^{acb}$, and there is a cancellation between the real parts of the $\cal{T}$ integrands, i.e. $K_{221}$ is a real quantity, symmetric in the exchange of all variables of gluons 1 and 2 (momentum, Lorentz, colour). One sees explicitely that the existence of the amplitudes with two indices 2 is related to the presence of the damping caused by $\hat{C}$. For $\hat{C}\ll P_i$, $K_{221}$ is proportional to $if^{acb}(i\hat{C})$. Ward identities connect these functions to the 2-point ones, in particular (the colour  is factored out) \cite{FG4}
\begin{eqnarray}
-i P_{1\mu}K_{221}^{\mu\nu\rho}(P_1,P_2,P_3) &=& K_{22}^{\nu\rho}(P_2) \nonumber \\
-i P_{3\rho}K_{221}^{\mu\nu\rho}(P_1,P_2,P_3) &=& K_{22}^{\mu\nu}(P_1)-K_{22}^{\mu\nu}(P_2)
\end{eqnarray}   
An example in position space is the case $\mu=\nu=\rho=0$
\begin{eqnarray}
\lefteqn{K_{221}^{000}(X,Y,Z)=- i f^{acb} T  \int {d_4P_1\over(2\pi)^4} {d_4P_2\over(2\pi)^4} \  {\cal{T}}^{000}(P_1,P_3,P_2) \nonumber} \\  & &
\sin(P_1+P_2).({-Z +(X+Y)/2}) \  \sin(P_1-P_2).{(X-Y)/2}
\label{b9}
\end{eqnarray}
antisymmetric in the exchange $X \leftrightarrow Y$ 

One now deduces from Eqs.(\ref{b8b})(\ref{b8}) the expressions for the connected Green functions. The needed relations were written in Sec.\ref{sec21}, i.e. Eqs.(\ref{a3b})(\ref{a5})(\ref{a6}), together with the approximation $N(P_i)\approx T/p_i^0$. One obtains
\begin{equation}
G_{211}^c=G_r(P_1)G_a(P_2)G_a(P_3) \ K_{211}
\end{equation}
\begin{eqnarray}
G_{221}^c=G_r(P_1)G_a(P_2)G_a(P_3) \ {\cal{T}}(1,3,2) \ T \ if^{acb} \nonumber \\  
+ G_a(P_1)G_r(P_2)G_a(P_3) \ {\cal{T}}(2,3,1) \ T\ if^{bca} \nonumber \\ 
+ G_r(P_1)G_c(P_2)G_a(P_3) \ {\cal{T}}(1,2,3) \ (-p_3^0)\  if^{abc} \nonumber \\ 
+G_c(P_1)G_r(P_2)G_a(P_3) \ {\cal{T}}(2,1,3) \ (-p_3^0)\ if^{bac} 
\end{eqnarray}
One notices that the following product always appear 
$$ G_r(P_i) \  {\cal{T}}(i,j,k) \ G_a(P_k).$$ 
Compared to $K_{221}$, i.e. Eq.(\ref{b8}), $G_{221}$ has two extra $\cal{T}$ terms, each one multiplied by a factor $G_c(P)$. All terms are of the same order as $G_c(P)$ carries a factor $T/p_0$. Endly, from Eq.(\ref{a6})
\begin{eqnarray}
\lefteqn{G_{222}^c=[ \ G_r(P_1) G_c(P_2) G_c(P_3) K_{211} + \ (1\leftrightarrow 2) + (1 \leftrightarrow 3) \ ] \nonumber } \\ & &
+  [ \ G_r(P_1) G_r(P_2) G_c(P_3) K_{221} + \ (3 \leftrightarrow 2) + (3 \leftrightarrow 1) \ ] 
\label{b10b}
\end{eqnarray}
\begin{eqnarray}
\lefteqn{G_{222}^c= T \  G_c(P_3) \ [ \  G_r(P_1)G_a(P_2) {\cal{T}}(1,3,2)  \nonumber} \\  & & 
-G_a(P_1)G_r(P_2) {\cal{T}}(2,3,1) \ ] \ if^{acb} \ + (3 \leftrightarrow 1) + (3 \leftrightarrow 2)
\label{b10}
\end{eqnarray}
where $(3 \leftrightarrow 1)$ means a term obtained by the exchange of all variables of gluons 3 and 1 (the colour order follows the time-order).   One goes from Eq.(\ref{b10b}) to  Eq.(\ref{b10}) by noting that the quantity ${\cal T}(1,3,2)$ occurs in $K_{211}$ and in $K_{221}$ so that
\begin{equation}
-p_2^0 \  G_c(P_2) +T \ G_r(P_2) = T \ G_a(P_2)   \label{b10c}
\end{equation}
    Alternatively, $G_{222}^c$ may be written in terms of the retarded functions (see Appendix \ref{App1}). One sees that while the 1PI amplitude $K_{222}=0$, the amplitude $G_{222}^c$ differs from zero. It is completely symmetric in the exchange of all variables of any pair of gluons. For $\hat{C}\ll P_i$, $G_{222}^c$ is proportional to $if^{abc}(i\hat{C})^2$. \\
One may convince oneself that the antisymmetry of the colour coefficient $f^{abc}$ does not force the symmetric $G_{222}^c(X,Y,Z)$ to vanish. A simpler symmetric example is the sum $(K_{221}+K_{212}+K_{122})(X,Y,Z)$ for the case $\mu=\nu=\rho=0$, with $K_{221}^{000}$ given by Eq.(\ref{b9}). This combination is antisymmetric in the exchange of any two space-time variables, it vanishes when two point coincide, or at mid-points such as $Z=(X+Y)/2$, however it does not vanish in general. As it has been argued in Sec.\ref{sec23}, $G_{222}(X,Y,Z)$ only involves anticommutators of the field $A_{\mu}$, and it should correspond to the 3-point correlation function of statistical physics. Then,  $G_{222}^c(X,Y,Z)$ is the third cumulant of these correlation functions. As it does not vanish, one concludes that the fluctuations of the $A_{\mu}$ field do not have a gaussian character. \\

One now  considers the 4-point function $G_{2222}^c$. The terms that come from the 4-point 1PI vertex have a structure similar to the 3-point one, i.e. Eq.(\ref{b10}), through the same process (\ref{b10c})
\begin{eqnarray}
\lefteqn{G_{2222}^{c \ (1)}= T \  G_c(P_2) G_c(P_3) \ [ \  G_r(P_1)G_a(P_4) {\cal{T}}(1,2,3,4)  \nonumber} \\  & &
+G_a(P_1)G_r(P_4) {\cal{T}}(4,3,2,1)   \ + (2 \leftrightarrow 3)\ ] \ +  \ 5 \  {\mathrm{terms}}
\label{b16}
\end{eqnarray}
where + 5 terms means the five other ways of choosing a pair among $(1,2,3,4)$, and 
\begin{eqnarray} 
\lefteqn{{\cal{T}}(1,2,3,4)= f^{abm}f^{mcd}\int_{all \ v} \nonumber} \\ & &
 v_{\mu}(v.P_1+i\hat{C})^{-1} v_{\nu}'(v.(P_1+P_2)+i\hat{C})^{-1}v_{\rho}''(-v.P_4+i\hat{C})^{-1}v_{\sigma}'''
\label{b16b}
\end{eqnarray}
Ward identities relate ${\cal{T}}(1,2,3,4)$ to ${\cal{T}}(i,j,k)$. For $\hat{C}\ll P_i$,  $G_{2222}^{c \ (1)}$ is proportional to $(i\hat{C})^3$. The contribution  $G_{2222}^{c \ (2)}$ from a pair of 3-point vertices joined by one dressed propagator is obtained from Eq.(\ref{a9}). The resulting expression is lengthy, there are terms with $G_r(P_i+P_j), G_a(P_i+P_j)$ and $G_c(P_i+P_j)$. For $\hat{C}\ll P_i$, it is proportional to  $(i\hat{C})^3$. $G_{2222}^{c \ (2)}$ 's structure is different from $G_{2222}^{c \ (1)}$ 's one and the sum does not vanish (they cooperate in current conservation in some circumstances). One concludes that the cumulant of order 4 of the correlation functions does not vanish. \\
For the $n$-point case, the term that comes from the $n$-point 1PI vertex is an immediate generalization of the 3- and 4-point cases, i.e. Eqs.(\ref{b10})(\ref{b16}). To conclude, the ultrasoft gluons exhibit an infinite series of correlation functions' cumulants $G_{22...2}^c$. These cumulants characterize the fluctuations that maintain the effective quantum system in equilibrium, at an intermediate scale between $(gT)^{-1}$ and $(g^2T)^{-1}$. 

\subsection{The HTL case}

The  HTL theory is the effective theory for the soft modes $p\ll T$ , i.e. when the scale $T^{-1}$  is integrated out. Here, the damping only occurs in the kinematical domain of space-like gluons. This case is a limiting case of the above formalism, as it corresponds to $p\sim gT  \gg \hat{C}\sim g^2T\ln 1/g$ , so that $\hat{C}$ may be approximated as
\begin{equation}
  C({\mathbf{v}, \mathbf{v'}}) \approx \epsilon \ \delta_{S_ 2}( {\mathbf{v}} -{\mathbf{v'}})  \ \ \ , \ \ \epsilon>0
\label{b11}
\end{equation}  
\begin{equation}
(v.P+i\hat{C})^{-1}\approx {\mathrm{P.P.}}{1 \over v.P} -i\pi \ \delta(v.P)
\label{b12}
\end{equation}
This case allows more explicit expressions for the amplitudes related to dissipation.
For the 2-point function, the result is well known as Landau damping \cite{MLB}. From Eq.(\ref{b5}) with (\ref{b12}), one obtains for $P^2=p_o^2-p^2>0$, for example
\begin{equation}
K_{21}^{i0}(P)=m_D^2 ({p_0 \over p})^2 p_i \  L(p_0,p) = m_D^2({p_0 \over p})^2 p_i \ {1\over 2p} \ln {\frac{p_o-p}{p_o+p}}
\label{b13}
\end{equation}
and from Eq.(\ref{b6}) with (\ref{b12})
\begin{equation}
K_{22}^{\mu\nu}={T\over p_0}(\Pi_r-\Pi_a) = m_D^2 T \ \int_v (-2i\pi)\   \delta(v.P) \ v_{\mu}v_{\nu}
\end{equation}    
For example
\begin{equation}
K_{22}^{00} =-m_D^2i\pi {T\over p}\  \theta(-P^2) \ \ \ , \ \ \ K_{22}^{i0} = {p_0p_i\over p^2} \ K_{22}^{00}
\end{equation}
consistent with $P_{\mu}K_{22}^{\mu\nu}=0$.  Moreover, $G_c=(G_r-G_a)T/p_0$ is, up to a factor, the transverse or longitudinal spectral function of the soft gluon \cite{MLB}. \\

For the 3-point function, 
${\cal{T}}^{\mu\nu\rho}(1,3,2)$ is given by Eq.(\ref{b7}) with the approximations (\ref{b11})(\ref{b12}). It has been explicited in Ref.\cite{FTay}. For the case $\mu=\nu=\rho=0$ and for $P_1^2>0 , P_2^2>0$
\begin{equation}
- m_D^{-2} \ {\cal{T}}^{000}(1,3,2) = M(P_1,P_2) = {1\over 2\sqrt{\Delta}} \ \ln{\frac{P_1.P_2 + \sqrt{\Delta}}{P_1.P_2 - \sqrt{\Delta}}}
\end{equation}
\begin{equation}
\Delta = (P_1.P_2)^2-P_1^2P_2^2 =\Delta(P_1,P_2)
\label{b14}
\end{equation}
$M(P_1,P_2)$ has branch points at $P_1^2=0, P_2^2=0$ and $\Delta=0$. At the branch point $P_1^2=0$, $M$ gets an imaginary part $\mp i\pi / 2\sqrt{\Delta}$.  \\
For space-like indices $\mu, \nu, \rho$, the tensor structure is decomposed on the vectors 
${\mathbf n}= {\mathbf p}_1\times {\mathbf p}_2 \ , \  {\mathbf n}\times {\mathbf p}_1 \ ,  \  {\mathbf n}\times {\mathbf p}_2 $ and one obtains the result that for $v_i$, or $v_iv_j$, or $v_iv_jv_k$, the integrals are linear functions of $M(P_1,P_2)$ and of $ L(P_1) ,  L(P_2)$ (defined in Eq.(\ref{b13})) with rational coefficients \cite{FTay}. \\
$K_{221}$ carries a definite prescription for the imaginary parts. From Eq.(\ref{b7}) with (\ref{b12})
\begin{equation}
K_{221}^{000}= m_D^2 T i f^{acb} { i \pi\over \sqrt{\Delta(P_1,P_2)}} [ \ \theta(-P_1^2) - \theta(-P_2^2) \ ]
\end{equation}  
with $\Delta$ defined in Eq.(\ref{b14}). For $K_{221}^{000}(P_1,P_2,P_3)$ to differ from zero, one needs either $P_1^2>0 ,  P_2^2<0 $ or $P_1^2<0  , P_2^2>0 $. 
\begin{eqnarray}
\lefteqn{K_{221}^{i00}={1\over {\mathbf n}^2}[\ p_2^0( {\mathbf n}\times {\mathbf p}_1)_i- p_1^0({\mathbf n}\times {\mathbf p}_2)_i \  ] K_{221}^{000} \nonumber}\\ & &
+m_D^2T {f^{acb}\over {\mathbf n}^2} [ \ ( {\mathbf n}\times {\mathbf p}_1)_i{\pi\over p_1}\theta(-P_1^2) + ({\mathbf n}\times {\mathbf p}_2)_i{\pi\over p_2}\theta(-P_2^2) \ ]
\end{eqnarray}
consistent with $$-i[p_1^0K_{221}^{000}-p_1^iK_{221}^{i00} ] =K_{22}^{00}(P_2).$$
Turning to $G_{222}^c$ , the term explicitely written in Eq.(\ref{b10}) differs from zero if $P_3^2<0$ and $P_1^2<0$ (or $P_2^2<0$). For example, for the case $\mu=\nu=\rho=0$, $G_{222}^c$ differs from zero in narrow domains of the plane $p_1^0+p_2^0+p_3^0=0$, such as the domain $ |p_1^0|<p_1, |p_2^0|>p_2$ with $|p_3^0|<(p_1^2+p_2^2+2 {\mathbf p}_1.{\mathbf p}_2)^{1/2}$. \\ 

A similar treatment may be made for the 4-point function. ${\cal{T}}(1,2,3,4)$ of Eq.(\ref{b16b}) may be expressed in terms of $M(P_i,P_j)$ and $L(P_k)$ functions \cite{FTay}. Its imaginary part may be obtained from Eq.(\ref{b16b}) with Eq.(\ref{b12}), and one sees that it is nonvanishing for $P_1^2<0$, or $P_4^2<0$, or $(P_1+P_2)^2<0$.  One concludes that the soft gluons exhibit the same type of correlations as the ultrasoft gluons.

\section{Classical correlation functions in ultrasoft QCD}
\label{sec4}
In this section, a comparison is made, at the scale $p\sim g^2T\ln1/g$, between the Green functions $G_{22..2}^c$ of the quantum approach and the corresponding quantities of the classical approach to ultrasoft QCD.
 D.B\"{o}deker's  method was to integrate out the scale $(gT)^{-1}$ in the kinetic equations associated with the field modes $p\ll T$.  The field modes $gT<p<T$ were split from the softer modes, their equation of motion was solved and plugged into the equation for the softer modes. A thermal average was performed over the initial conditions which enter that solution. The resulting classical equations for the ultrasoft field modes $p\ll gT$ are expressed in terms of an auxiliary colour field $W(X,{\mathbf v})$
\begin{equation}
(v.D_x+\hat{C}) W(X,{\mathbf v}) = {\mathbf v}.{\mathbf E}(X) + \xi(X,{\mathbf v})    \label{c1}
\end{equation}
\begin{equation}
[D_{\mu} , F^{\mu \nu}] (X) = m_D^2 \int_v \ v^{\nu} W(X,{\mathbf v})  \label{c2}
\end{equation}  
where $\xi(X,{\mathbf v})$ is a gaussian white noise with vanishing average, and with correlation function
\begin{equation}
\langle\langle \  \xi_a(X,{\mathbf v}) \ \xi_b(X',{\mathbf v}') \  \rangle \rangle = {2T\over m_D^2} C( {\mathbf v},{\mathbf v}') \ \delta_{ab} \ \delta^4(X-X')
\label{c3}
\end{equation}  
where $\langle \langle \ \ \ \ \rangle \rangle $ means an average over the random noise.  The noise arises from the average over initial conditions for the modes $gT<p<T$ that are integrated out. This noise maintains the ultrasoft modes in thermal equilibrium, as it compensates for the dissipation  due to $\hat{C}$. 

A related idea has been presented for the HTL theory, i.e. for the field modes $p\ll T$ \cite{AY3,Ia}. The classical equations for this $A_{\mu}$ field are Eqs.(\ref{c1})(\ref{c2}) with $\hat{C}=0$ and $\xi(X,{\mathbf v})=0$. A summary of the results are: (i) in the abelian case (or the abelian approximation to QCD) the average over the initial conditions of the $W$ field produces a fluctuating piece $\zeta(X)$ in the equation for the $A_i$ field (in the gauge $A_0=0$),  (ii) the average over the $W$ field's initial conditions amounts to a gaussian functional integral,  (iii) the fluctuating  $\zeta(X)$ generates the Landau damping part of the  $A_i$ field's 2-point function.   \\

 In this section, we solve the set of equations (\ref{c1})(\ref{c2}) as an expansion in $\xi$, and we compare the noise average of the 2-point and 3-point functions of the $A_\mu$ field with the correlation functions $G_{22}$ and $G_{222}$ of the quantum approach (see Sec.\ref{sec23}).  

Eq.(\ref{c1}) is solved by means of the Green function defined in Eq.(\ref{b1}). With initial condition $W=0$ at $t=-\infty$,
\begin{equation}
W(X,{\mathbf v}) =  i\int_Y \int_{v'} G_{ret}(X,Y;A;{\mathbf v},{\mathbf v}') \ [ \ {\mathbf v}'.{\mathbf E}(Y) + \xi(Y,{\mathbf v}') \ ] 
\end{equation}
with $\int_Y =\int d_4Y$ and $\int_{v'}=\int {d \Omega_{v'} / 4 \pi}$. Then, Eq.(\ref{c2}) for the ultrasoft field turns into
\begin{equation}
[D_{\mu} , F^{\mu \nu}]_a (X)=j_a^\nu(X,A) + j^\nu_{\xi, a}(X,A)   \label{c4}
\end{equation}
where $j_a^\nu(X,A)$ may be written as in Eq.(\ref{b3}) and 
\begin{equation}
j^\nu_{\xi, a}(X,A) = m_D^2  i\int_Y \int_{v,v'} v^\nu \ G_{ret}^{a b}(X,Y;A;{\mathbf v},{\mathbf v}') \ \xi(Y,{\mathbf v}')   \label{c5}
\end{equation}
Eq.(\ref{c4}) may be solved by iteration as an expansion in $\xi$, at the scale $X\sim (g^2T\ln 1/g)^{-1}$. First, one expands the currrents $j^\nu(X,A)$ and $j^\nu_{\xi, a}(X,A)$ in powers of the gauge field. This expansion is valid, i.e. $\partial_x \sim g^2T\ln1/g  \gg  gA \sim g^2T$    provided $\ln 1/g \gg 1$. From Eq.(\ref{b3c}), formally 
\begin{equation}
j^\mu(X,A) = \Pi^{\mu\nu} \ A_\nu + {g\over 2}\  \Gamma^{\mu\nu\rho} \ A_\nu \ A_\rho + \cdots
\label{c6}
\end{equation} 
and a similar one for $j^\nu_{\xi, a}(X,A)$.  One, then, writes Eq.(\ref{c4}) as 
\begin{equation}
(\partial^2g^{\mu\nu}-\partial^\mu\partial^\nu - \Pi^{\mu\nu}) \ A_{\nu} =   
   j^\mu_{\xi}(X,A)  -g F^{\mu\nu} A_{\nu} + {g\over 2} \Gamma^{\mu\nu\rho} A_\nu  A_\rho + \cdots
\label{c7}
\end{equation} 
To lowest order, one sets $A=0$ in the right-hand-side of Eq.(\ref{c7}). The retarded dressed propagator $G_r^{\mu\nu}(X,Y)$ is an inverse of $(\partial^2g_{\mu\nu}-\partial^\mu\partial^\nu + \Pi^{\mu\nu})$ , and the lowest order solution to Eq.(\ref{c7}) is (with initial condition $A=0$ at $t=-\infty$)
\begin{equation}
A^{(0) \mu}_a(X) = G_r^{\mu\mu'}(X,X_1) \ m_D^2 i \  v_{1\mu'} \ {\cal G}_r(X_1,X_2; {\mathbf v}_1,{\mathbf v}_2) \ \xi_a(X_2,{\mathbf v}_2)
\label{c8}
\end{equation}  
with the convention that a repeated space-time, or ${\mathbf v}$,  variable means a summation over this variable, and with the definition
\begin{equation}
{\cal G}_r(X_1,X_2;{\mathbf v}_1,{\mathbf v}_2) = G_{ret}(X_1,X_2; A=0 ; {\mathbf v}_1,{\mathbf v}_2)
\label{c9}
\end{equation}   
 One now considers the average over the random noise. $\langle\langle \ A^{(0)\mu}_a(X) \  \rangle\rangle =0$ from $\langle\langle \ \xi \ \rangle\rangle =0$. The computation of   $\langle\langle \ A^{(0)\mu}_a(X) A^{(0)\nu}_b(Y) \ \rangle\rangle$ is instructive, and the steps are detailed in Appendix \ref{App2}. The result is
\begin{equation}
 \langle\langle \ A^{(0)\mu}_a(X) A^{(0)\nu}_b(Y) \ \rangle\rangle =\int {d_4P_1\over (4\pi)^4} \   e^{-iP_1.(X-Y)} \delta_{ab}  \  i  G_c^{\mu\nu}(P_1)   \label{c10}
\end{equation}
with $G_c(P)$ defined in Eq.(\ref{b6b}). Since $G_c(P)=G_c(-P)$, this 2-point function is  symmetric in the exchange $X\leftrightarrow Y$. 

A result equivalent to (\ref{c10}) was obtained in the original paper on ultrasoft \cite{Bod1} in the static case, and also for the HTL case \cite{Ia, AY3}. One may compare the expression (\ref{c10}) with the quantum theory result, Eqs.(\ref{a12})(\ref{b6b}), and one sees that
\begin{equation}
-i \langle\langle \ A^{(0)\mu}_a(X) A^{(0)\nu}_b(Y) \ \rangle\rangle =G_{22}^c(X,Y)=-{i\over 2}\langle \  \lbrace A^{(0)\mu}_a(X) \ , \ A^{(0)\nu}_b(Y) \rbrace\  \rangle 
\end{equation}
One finds an exact correspondence between the (noise-averaged) classical correlation function and the matrix element of the  field operators' anticommutator.  All the higher order cumulants of $A^{(0)\mu}_a(X)$ vanish, since, from Eq.(\ref{c8}), $ A^{(0)\mu}_a(X)$ is linear in the gaussian noise $\xi$. \\  

Turning to the next iterated solution  to Eq.(\ref{c7}), $ A^{(0)\mu}_a(X)+ A^{(1)\mu}_a(X)$,  it is obtained with the substitution of $ A^{(0)\mu}_a(X)$ to $ A^{\mu}_a(X)$ in the following three terms of the r.h.s. of Eq.(\ref{c7}): the terms involving the bare 3-point vertex, the effective vertex $\Gamma^{\mu\nu\rho}$, and the second term in the expansion of $j^\mu_{\xi, a}(X,A)$. These terms are, then,  bilinear in $\xi$ and such that
\begin{eqnarray}
\langle\langle \ A^{(1)\mu}_a(X) \  \rangle\rangle &=& 0 \ \ \ {\mathrm{since}}\ \  f_{abc} \delta_{bc} =0   \\
\langle\langle \ A^{(1)\mu}_a(X)  A^{(0)\nu}_b(Y)\  \rangle\rangle &=& 0  \ \ \ {\mathrm{since}}\ \ \langle\langle \ \xi\xi\xi \  \rangle\rangle=0
\end{eqnarray}
The next step is to compute $\langle\langle \ A^{(1)\mu}_a(X)A^{(0)\nu}_b(Y)A^{(0)\rho}_c(Z) \  \rangle\rangle$ and to compare the symmetric quantity
\begin{equation}
G_3 = \langle\langle \ A^{(1)\mu}_a(X)A^{(0)\nu}_b(Y)A^{(0)\rho}_c(Z) \  \rangle\rangle +(X,\mu,a \leftrightarrow Y,\nu,b) + (X,\mu,a\leftrightarrow Z,\rho,c)
\label{c12}
\end{equation}   
with the quantum result $G_{222}^c$. One defines the Fourier transform
\begin{eqnarray}
\lefteqn{\langle\langle \ A^{(1)\mu}_a(X)A^{(0)\mu}_b(Y)A^{(0)\mu}_c(Z) \  \rangle\rangle = \int \prod_{i=1}^3{d_4P_i\over(2\pi)^4} \  e^{-i(P_1.X+P_2.Y+P_3.Z)}  \nonumber} \\ & &
(2\pi)^4\delta^4 (P_1+P_2+P_3) \ {\cal A}^{\mu\nu\rho}_{abc}(P_1,P_2,P_3)
\label{c13} 
\end{eqnarray}   
One considers successively the contribution of the three terms. With the bare vertex contribution
\begin{equation}
A^{(1,\alpha)\mu}_a(X) =-G_r^{\mu\mu'}(X,X_1) \ g f_{abc} 
 \ A^{(0)\sigma}_b(X_1) \ [ \ \partial_{\mu'}A^{(0) c}_\sigma - \partial_{\sigma} A^{(0) c}_{\mu'}(X) \ ](X_1)
\label{c14}
\end{equation}   
one obtains (see Eq.(\ref{Ap6}) in Appendix \ref{App2})
\begin{eqnarray}
\lefteqn{{\cal A}^{\mu\nu\rho \ (\alpha)}_{abc}(P_1,P_2,P_3)=g \ G_r^{\mu\mu'}(P_1) \lbrace  \nonumber} \\ & & 
 i f_{abc} \ G_c^{\nu\sigma}(P_2) \ [ \ P_{3\sigma}G^\rho_{c\  \mu'}(P_3) - \ P_{3\mu'}G^\rho_{c\  \sigma}(P_3) \ ] \ +  \ \ (P_2,\nu,b \leftrightarrow P_3,\rho,c)   \ \rbrace
\label{c17}
\end{eqnarray}
With the effective vertex contribution
\begin{equation}
A^{(1,\beta)\mu}_a(X) =G_r^{\mu\mu'}(X,X_1)  \ {g\over 2} \Gamma^{abc}_{\mu'\nu'\rho"}(X_1,X_2,X_3) \ A_b^{(0)\nu}(X_2)A_c^{(0)\rho}(X_3)
\label{c18}
\end{equation}
the result is 
\begin{equation}
{\cal A}_{abc}^{\mu\nu\rho \ (\beta)}=-G_r^{\mu\mu'}(P_1)  \lbrace  \ G_c^{\nu\nu'}(P_2) G_c^{\rho\rho'}(P_3)   
 {g\over 2}\Gamma^{abc}_{\mu'\nu'\rho'}(P_1,P_2,P_3) \ \ + \ (\ P_2,\nu,b \leftrightarrow P_3,\rho,c \ ) \ 
\rbrace
\label{c19}
\end{equation}
with the correspondence between the notation in the $R/A$ basis and in the Keldysh bases $\Gamma_{RAA}(P_1,P_2,P_3) = K_{211}(P_1,P_2,P_3)$, and with the explicit expression (\ref{b8b})(\ref{b7}) for $K_{211}$. The symmetry of $K_{211}$ in the exchange of the variables 2 and 3 leads to an identical term in Eq.(\ref{c19}) so that $2 \times g/2 =g$.  \\
Comparing the contributions $(\alpha)$ and $(\beta)$, i.e. Eqs.(\ref{c17}) and (\ref{c19}), one sees that the bare vertex' one is subleading for momenta $ p\sim g^2T \ln 1/g $, as the effective vertex carries an extra factor of order $m_D^2/(v.P)^2$.  One now compares Eq.(\ref{c19}) with the form (\ref{b10b}) for the quantum correlation function $G_{222}^c$ , and one sees that ${\cal A}^{(\beta)}(P_1,P_2,P_3)$ is the term $G_r(P_1)G_c(P_2)G_c(P_3) K_{211}$, so that the symmetric combination defined in Eq.(\ref{c12}) corresponds to the three terms, in the form (\ref{b10b}), that have one Keldysh index 2. (In Sec.\ref{sec3}, $g$ has been factored out of the vertex, see Eq.(\ref{b3c}))  

Endly, the term arising from the second term in the expansion of $j^{\mu}_\xi(X,A)$ is
(see Eq.(\ref{b3d}))
\begin{eqnarray}
\lefteqn{A_a^{(1,\gamma) \mu}(X)=G_r^{\mu\mu'}(X,X_1) \ m_D^2\ i g  \ i f_{adg} \ A_d^{(0)\nu'}(X_2) \nonumber} \\ & & 
\times \ \ v_{1\mu'} {\cal G}(X_1,X_2,{\mathbf v}_1 ,{\mathbf v}_2) v_{2\nu'} {\cal G}(X_2,X_3,{\mathbf v}_2,{\mathbf v}_3) \ \xi_g(X_3,{\mathbf v}_3)
\label{c20}
\end{eqnarray}   
After some computation, one obtains (see Appendix \ref{App2}  and Eq.(\ref{Ap7}))
\begin{equation}
{\cal A}_{abc}^{\mu\nu\rho (\gamma)} = - G_r^{\mu\mu'}(P_1) \lbrace  \ G_c^{\nu\nu'}(P_2) G_r^{\rho\rho'}(P_3)   
 \ g\ i f_{abc} \ [ \ {\cal T}(1,2,3) + S \ ] \ + \  \ (P_2,\nu, b \leftrightarrow P_3,\rho, c)  \ \rbrace
\end{equation}
with ${\cal T}(i,j,k)$ defined in Eq.(\ref{b7}), and where $S$ is a term that will disappear in the symmetric combination (\ref{c12}) (see Appendix \ref{App2}).  One, then,  sees that the symmetric combination of ${\cal A}^{(\gamma)}$ terms builds up the six  ${\cal T}(i,j,k)$ terms, as they are present in the sum of the terms with two Keldysh indices 2 of the form (\ref{b10b}). This comparison exhibits the correspondence  between the terms arising from the noise in the classical equation and the terms  related to the dissipation in the quantum approach. \\

To conclude, the symmetric combination (\ref{c12}) of noise-averaged approximate classical solutions lead to the same value as the matrix element of the combination of anticommutators defined in Eq.(\ref{a15}), as suggested in Sec.\ref{sec23}. This result brings comfort as, somehow, it was expected. Indeed, from the quantum theory point of view, the lack of freedom at the level of the 3-point functions is the same as for the 2-point ones. This is best seen in the $R/A$ basis where the set of 3-point functions is made of the three retarded ones, such as $\Gamma_{RAA}$, and their complex conjugate, the advanced ones. As a result, $G_{222}^c$ is, in fact, determined by the response functions, just as $G_{22}^c$ is (for the general relation,  see Appendix \ref{App1}).

We have not pushed  any further the comparison  between the classical and quantum results, due to the proliferation of terms in the classical approach. One may trace the terms that are present in $G_{2222}^{c \ (1)}$. The third term in the expansion of $j^{\mu}(X,A)$ will lead, in the 4-point correlation function, to the terms of the type $K_{2111}$, and the terms of the type $K_{2211}$ appear in the same expansion of  $j^{\mu}_{\xi}(X,A)$.

\section{Conclusion}
\label{sec5}
 We have followed a  quantum field theory approach to ultrasoft QCD. From the knowledge of the full set of $n$-point Green functions, one is able to extract information on how the quantum system compensates for the dissipation due to collisions.  
We have focused on the $n$-point functions which are likely to survive in the classical limit, i.e. the matrix elements of anticommutators of the gauge field operators. Their connected part should correspond to the cumulants of the correlation functions in statistical physics. This is one way to characterize the fluctuations that maintain the system in thermal equilibrium. \\
The result that has been obtained is that the whole series of cumulants differs from  zero in ultrasoft (or soft) QCD. This result suggests that there may be alternative ways to characterize these large-scale, classical fluctuations of the (non-abelian) gauge field.  \\
The third order cumulant has been given special attention. Explicit expressions have been  written down  for the ultrasoft effective theory ($p \ll gT$) and for the soft theory ($p \ll T$). Comparison has been made with the semiclassical approach to the effective theory, where a random noise compensates for the dissipation. The solution to the classical equation has been written as an expansion in powers of the noise, and the noise average of the 3-point function has been computed to lowest (non-zero) order. It agrees with the quantum result, which shows that the correspondence is correct between the quantum quantity and the classical one. Those results and comparisons have been made at the scale $(g^2T\ln 1/g)^{-1}$, assuming $\ln1/g \gg 1$, i.e. at an intermediate scale between $(gT)^{-1}$ and  $ (g^2T)^{-1}$. This scale allows an expansion, in powers of the gauge field, of the Green function $(v.D+i\hat{C})^{-1}$ so that  explicit expressions have been obtained and compared.\\

An extensive use has been made of the special features of the Keldysh basis. The properties of the ultrasoft amplitudes may be summarized in the following term in an effective action
\begin{equation}
\Gamma^{pot}(A_1,A_2)={\mathrm{Tr}}[\  \int_X A_2^\mu(X) \ j_\mu(X; A_1(X)  )  \\
+ \int_{X,Y} {1\over2} \ A_2^\mu(X) \ {\cal K}_{\mu \nu}(X,Y; A_1(X)  ) \ A_2^\nu(Y)  \ ] \label{d1}
\end{equation}
with $\int_X=\int_{-\infty}^{+\infty}dt \int d_3x$. 
$A_1(X)$ and $A_2(X)$ are the two components of the gauge field in the Keldysh basis, $j_\mu(X;A)$ and ${\cal K}_{\mu \nu}(X,Y; A)$  are non-local quantities expressed in terms of the propagator (along jagged paths) in the background non-abelian field $A$. Their expressions may be found in  Eqs.(\ref{b3})(\ref{b3b}). ${\cal K}_{\mu\nu}$ would vanish if there were no dissipation.  Eq.(\ref{d1}) explicitly shows  that the only existing 1PI ultrasoft vertices  have either one or two Keldysh indices 2. The Keldysh basis is such that 
\begin{equation}
A_1(X)=(A_+(X)+A_-(X))/2         \ \ \ , \ \ \ \  A_2(X)= A_+(X)-A_-(X)
\end{equation}
where $A_+$ is the field that lives on the forward branch  of the time contour and $A_-$ on the backward branch (in the path integral formalism). Then $A_1(X)$ is the average $(A_++A_-)/2$  and  plays the role of a background field, whereas  $A_2(X)$ is the difference $ A_+-A_-$ and  plays the role of a fluctuating field. \\
This interpretation is made stronger if one considers  the  gauge transformation
\begin{equation}
A^\mu \rightarrow h\ A^\mu \ h^\dagger - (i / g) \  h \ \partial^\mu \ h^\dagger
\label{d2}
\end{equation}
In this transformation, the propagator in the background field $A$ transforms covariantly, as do $j_\mu(X ;A)$ and ${\cal K}_{\mu \nu}(X,Y ; A)$ ,  $j_\mu\rightarrow h j_\mu h^\dagger  $ ,  $ {\cal K}_{\mu\nu}\rightarrow h {\cal K}_{\mu\nu} h^\dagger$ \   ($D^\mu j_\mu=0= D^\mu_X {\cal K}_{\mu \nu}=D^\nu_Y {\cal K}_{\mu \nu}$) { \cite{BI4}.
With the identification $A_1^{\mu}(X)=A^\mu(X)$,  $\Gamma^{pot}(A_1,A_2)$ , defined in Eq.(\ref{d1}),  will be invariant under the transformation (\ref{d2})  if $A_2^{\mu}$ transforms covariantly 
\begin{equation}
A_2^{\mu}\rightarrow h \ A_2^{\mu}\  h^\dagger
\label{d3}
\end{equation}
The set of transformation rules (\ref{d2})(\ref{d3})  are those of the background field method. Here, the gauge field is split into a background field $A_\mu(X)$ (to be identified with the average field) and a fluctuating field $a_\mu(X)$ which transforms covariantly under the gauge transformation of the background field, ${\cal A}_\mu =A_\mu+a_\mu$. One concludes that the form (\ref{d1}) is in the class of  background field gauges \cite{BI4}.  It is worth stressing that the explicit form of the collision operator $C({\mathbf v} , {\mathbf v}')$ is irrelevant. Its required properties are: symmetry in ${\mathbf v}\leftrightarrow {\mathbf v}'$, positive eigenvalues,  the zero mode $\int_v C({\mathbf v} , {\mathbf v}') = 0$,  and gauge-fixing independence. \\

With the interpretation of the $A_2(X)$ field as a fluctuating field, one  is led to  the idea that these fluctuations are, very likely,  the ones  that compensate for the dissipation due to collisions and maintain  the system in thermal equilibrium.  An encouraging fact is that, for a classical field, 
\begin{equation}
-{i\over 2 T} \int_{X,Y} F_a^{0 \mu}(X) \ {\cal K}^{a b}_{\mu \nu}(X,Y;A) \ F_b^{\nu 0}(Y) 
 = {\mathrm{Sym.Part}}   \left( \int_X  F_a^{0 \mu}(X) \ j^a_\mu(X ; A)  =\int_X {\mathbf j}.{\mathbf E}(X) \right)
\end{equation}
 where Sym.Part means the symmetric part in the exchange $E_a^i(X)\leftrightarrow E_b^j(Y)$ , and 
$ {\mathbf j}.{\mathbf E}(X)$ is an energy density rate. 

\newpage
\appendix
\section{Constraint on n-point functions \\ and fluctuation-dissipation theorem} \label{App1}
In statistical physics, those theorems relate the fluctuations, i.e. a correlation function's cumulant, to the response of the system to an external weak perturbation, and, mostly, they involve 2-point functions. In field theory, the response functions are the $n$-point retarded amplitudes (an example is the current of Sec.\ref{sec3}) and the statistical correlations may be identified with $G_{22\cdots2}$ at high temperature (see Sec.\ref{sec23}). \\
There exists relations among the $2^n$ $n$-point amplitudes which are best stated as a complex conjugate relation in the $R/A$ basis. In addition to $\Gamma_{RR\cdots R}=0=\Gamma_{AA\cdots A}$ , one has, for a boson system, \cite{vWK,FG4}
\begin{eqnarray}
{\cal N}(P_1, \ldots P_k) \ \Gamma(P_{1A} \ldots P_{kA}, P_{(k+1)R} \ldots P_{nR}) = \nonumber \\
(-1)^n {\cal N}(P_{k+1} \ldots P_n) \ \Gamma^*(P_{1r} \ldots P_{kr}, P_{(k+1)A} \ldots P_{nA})
\label{Ap10}
\end{eqnarray} 
where
\begin{equation}
 {\cal N}(P_1, \ldots P_k) = \prod_{i=1}^k (N(P_i)+1/2) -\prod_{i=1}^k (N(P_i)-1/2)
\end{equation}
and $N(P_i)= (1/2) \coth \beta p_i^0/2$. In particular, ${\cal N}(P)=1$.(The $(-1)^n$ factor depends on conventions). \\

For the 2-point function, this relation is $\Gamma_{RA}=\Gamma_{AR}^*$ and, with $\Gamma_{RR}=0$, the relation  between the $R/A$ and Keldysh bases  written in Sec.\ref{sec2}, i.e. Eq.(\ref{a4}), leads to 
\begin{equation}
K_{22}(P)=N(P) [ \Gamma_{RA}-\Gamma_{AR}] = N(P)  \ [\  K_{21}-K_{21}^*\ ]
\end{equation}
and to a similar one between the correlation function $G_{22}^c$ and the response function $G_r$
\begin{equation}
G_{22}^c(P)= N(P) \ [ G_r-G_a](P)
\end{equation}
This is a well known fluctuation-dissipation theorem. \\

 For the case of the 3-point functions, there exists an equivalent relation. Indeed, Eq.(\ref{Ap10}) relates $\Gamma_{RRA}$ to $\Gamma_{AAR}^*$. The relation stemming from $\Gamma_{RRR}=0$ is \cite{FG4}
\begin{eqnarray}
-K_{222}=N(P_1)K_{122} + N(P_2)K_{212} + N(P_3)K_{221}  \nonumber \\
+ N(P_2) N(P_3) K_{211} + N(P_1) N(P_3) K_{121} + N(P_1) N(P_2) K_{112}
\end{eqnarray}
Now, Eq.(\ref{a4}) and Eq.(\ref{Ap10}) allow to express $K_{222}$ in terms of the retarded functions and their complex conjugate. The result is
\begin{eqnarray}
 K_{222}=  N(P_2) N(P_3) (K_{211}-K_{211}^* )+ N(P_1) N(P_3) (K_{121}-K_{121}^*) \nonumber \\ + N(P_1) N(P_2) (K_{112}-K_{112}^*)  \ -(K_{211}^*+K_{121}^*+K_{112}^*)/4
\label{Ap12}
\end{eqnarray}
where the following relation holds for $P_1+P_2+P_3=0$
\begin{equation}
N(P_1) N(P_2)+N(P_1) N(P_3)+N(P_2) N(P_3) +1/4 =0
\label{Ap13}
\end{equation}
$K_{222}$ is completely determined by the retarded 3-point functions, just as $K_{22}$ is in terms of the 2-point one. However, $K_{22}$ vanishes if there is no dissipation, while $K_{222}$ does not necessarily do so. For a bare vertex (i.e. no dissipation) $K_{222} \not=0$. Indeed, in this case,  the three retarded functions are identical, such that $K_{211}=-K_{211}^*$,  so that relation (\ref{Ap12}) becomes
\begin{equation}
K_{222}^{bare} =K_{211}^{bare} \ [-{1\over 4}\times2 +{1\over4}\times3] = {1\over4} K_{211}^{bare}
\end{equation}
in agreement with Ref.\cite{vWK} (where, for a bare $n$-point vertex, the following result is obtained: all bare vertices with an even number of 2 vanish, those with an odd number are proportional). However,  at high T, this bare vertex $K_{222}$  is lacking a factor $T/p_i^0$ and is subleading compared to the effective vertices. These vertices  obey   relation  (\ref{Ap12})  with  the approximation $N(P_i)\approx T/p_i^0$ ( and relation (\ref{Ap13}) turns into ${T^2\over {p_1^0p_2^0}}+{T^2\over {p_2^0p_3^0}}+{T^2\over {p_1^0p_3^0}}=0$). It remains to be seen whether this approximated form of Eq.(\ref{Ap12}) may be called, in general, a fluctuation-dissipation theorem, at high $T$, for the 3-point function.   

It is worth noticing that, by the argument that leads from Eq.(\ref{a3}) to Eq.(\ref{a3b}),  one deduces that the correlation function $G_{222}^c$ obeys a relation similar to Eq.(\ref{Ap12}) with the following substitutions: $G_{211}^c=G_r(P_1)G_a(P_2)G_a(P_3) \ K_{211}$ substituted to $K_{211}$ ,  \\ $G_{211}^{c\ *}=G_a(P_1)G_r(P_2)G_r(P_3) \ K_{211}^*$ substituted to $K_{211}^*$ , and so forth for the other terms of Eq.(\ref{Ap12}). This relation for $G_{222}^c$ was obtained in \cite{Ca3} from the examination of the 3-point spectral functions. \\

Turning to the 4-point case,  $\Gamma_{RRRA}$ is expressed in terms of $\Gamma_{AAAR}^*$ .  However, the relation between $\Gamma_{RRAA}$ and $\Gamma_{AARR}^*$ is only a constraint, and does not determine those amplitudes. Consequently, $K_{2222}$ cannot be expressed only in terms of the response functions, i.e. the functions with one Keldysh index 2. More generally, for an $n$-point amplitude in the $R/A$ basis, there are $2^n-2=2(2^{n-1}-1)$ non-vanishing amplitudes, with $2^{n-1}-1$ complex conjugate relations, and only $n$ response functions. \\
The $n$-gluon ultrasoft amplitudes are less numerous, because of Bose and gauge symmetries. There is one generating functional for the $n$ response functions (one Keldysh index 2), and another one for the $n(n-1)/2$ functions related to dissipation (two Keldysh indices 2). In the correlation functions (which describe the fluctuations) there is an interplay between the two sets of functions. The appearance of the same Green function in both generating functionals, may be seen as the ultimate fluctuation-dissipation constraint. \\
Coming back to the general case, one may tentatively identify the complex-conjugate relation (\ref{Ap10}) with the general constraint of the fluctuation-dissipation theorem.

\section{Noise average of classical solutions} \label{App2}
This appendix details the computation of the noise average of the classical 2-point and 3-point functions examined in Sec.\ref{sec4}. One goes to Fourier space. For the noise
\begin{equation}
\xi_a(X,{\mathbf v}) =\int {d_4P \over (4\pi)^4}  \ e^{-iP.X} \ \xi_a(P, v)
\end{equation}
and Eq.(\ref{c3}) translate into
\begin{equation}
\langle\langle \ \xi_a(P_1, v_1) \ \xi_b(P_2,v_2) \ \rangle\rangle = \  (2\pi)^4 \delta^4(P_1+P_2)\ \delta_{ab} \ {T\over m_D^2} \ 2 C( v_1, v_2)
\label{Ap1}
\end{equation}
For the lowest order solution, Eq.(\ref{c8}) translates into
\begin{equation}
A_a^{(0)\mu}(X)=\int {d_4P_1\over(2\pi)^4} \  e^{-iP_1.X} \ G_r^{\mu\mu'}(P_1) \ m_D^2 \  i \ v_{1\mu'}(v.P_1+i\hat{C})^{-1}_{v_1v_2} \ \xi_a(P_1,v_2)
\label{Ap1b}
\end{equation}
 with the convention that a repeated $v$ variable is summed over (compare with Eq.(\ref{b5})).  \\

One now details the computation of the 2-point function. From Eq.(\ref{Ap1b})
\begin{eqnarray} 
\lefteqn{\langle\langle \ A_a^{(0)\mu}(X) A_b^{(0)\nu}(Y) \ \rangle\rangle =(i)^2\int \prod_{i=1}^2{d_4P_i\over(2\pi)^4}  \ e^{-i(P_1.X+P_2.Y)} (2\pi)^4\delta^4(P_1+P_2)  \nonumber }\\ & & 
m_D^4\ G_r^{\mu\mu'}(P_1) G_r^{\nu\nu'}(P_2) \ 
\ v_{1\mu'}(v.P_1+i\hat{C})^{-1}_{v_1v_2} \ \delta_{ab} {T\over m_D^2} 2C(v_2,v_4) \ (v.P_2+i\hat{C})^{-1}_{v_4v_3} v_{3\nu'}
\label{Ap2}
\end{eqnarray}  
where Eq.(\ref{Ap1}) has been used. $C(v,v')$ is symmetric in the exchange $v\leftrightarrow  v'$, and so is $(v.P+i\hat{C})^{-1}_{vv'}$ . Then, with $ 2\hat{C} =[\  v.P_1+i\hat{C} + (-v.P_1+i\hat{C})\  ]\ /i$, one writes
\begin{equation}
(v.P_1+i\hat{C})^{-1} \ 2 \hat{C} \ (-v.P_1+i\hat{C})^{-1}= [ \ (v.P_1+i\hat{C})^{-1}+(-v.P_1+i\hat{C})^{-1}] /i
\label{Ap4}
\end{equation}
Comparing with Eq.(\ref{b6}), one recognizes a factor $K_{22 \ \mu'\nu'}(P_1)$ in Eq.(\ref{Ap2}). Moreover, since $G_r(P_2)=G_r(-P_1)=G_a(P_1)$,
\begin{equation}
G_r^{\mu\mu'}(P_1) \ K_{22 \ \mu'\nu'}(P_1) \ G_a^{\nu'\nu}(P_1) = {T\over p_1^0} \ G_r^{\mu\mu'}\ (\Pi_r-\Pi_a)_{\mu'\nu'} \ G_a^{\nu'\nu} =\ G_c^{\mu\nu}(P_1)
\end{equation} 
from Eq.(\ref{b6b}), so that Eq.(\ref{Ap2}) turns into
\begin{equation}
\langle\langle \ A_a^{(0)\mu}(X) A_b^{(0)\nu}(Y) \ \rangle\rangle =\int {d_4P_1\over(2\pi)^4}  \ e^{-iP_1.(X-Y)}  \ \delta_{ab} \  i  \ G_c^{\mu\nu}(P_1)
\label{Ap5}
\end{equation} \\ 

One now turns to the 3-point correlation function. The contribution coming from the bare 3-point vertex is obtained from Eq.(\ref{c14}) 
\begin{eqnarray}
\lefteqn{\langle\langle \ A_a^{(1,\alpha)\mu}(X) A_b^{(0)\nu}(Y) A_c^{(0)\rho}(Z) \ \rangle\rangle  = \  -G_r^{\mu\mu'}(X,X_1) \ g\ f_{adg} \nonumber} \\ & &
\lbrace   \ \langle\langle \ A_d^{(0)\sigma}(X_1) A_b^{(0)\nu}(Y) \ \rangle\rangle \ \ 
\langle\langle \ (\partial_{\mu'}A_\sigma^{(0) g} - \partial_{\sigma}A_{\mu'}^{(0) g})(X_1) A_c^{(0)\rho}(Z) \ \rangle\rangle \nonumber \\ & & 
+ \ ( Y,\nu, b \leftrightarrow Z,\rho,c ) \ \rbrace
\label{Ap6}
\end{eqnarray}   
With Eq.(\ref{Ap5}), one immediately obtains Eq.(\ref{c17}) in momentum space. A similar treatment applies to the effective vertex contribution.  

For the third contribution, $A^{(1,\gamma)}(X)$ is in Eq.(\ref{c20}), and one needs, in addition to Eq.(\ref{Ap5}), another noise average. From Eqs.(\ref{c8})(\ref{c3})
\begin{equation}
 \langle\langle \  \xi_g(X_3,v_3) \ A_c^{(0)\rho}(Z) \ \rangle\rangle = T \ \delta_{gc} G_r^{\rho\rho'}(Z,X_4)\   i \ v_{4\rho'} {\cal G}_r(X_4,X_3;v_4,v_5) \ 2C(v_5,v_3) 
\end{equation}
with the same conventions as in Eq.(\ref{c8}). The 3-point  correlation function is, then, in momentum space and with the definition (\ref{c13}),
\begin{eqnarray}
\lefteqn{{\cal A}_{abc}^{\mu\nu\rho \ (\gamma)} =-G_r^{\mu\mu'}(P_1) \ \lbrace \ G_r^{\rho\rho'}(P_3) \   i G_c^{\nu\nu'}(-P_2) \ g \ m_D^2 \ T \ i f_{abc}  \nonumber} \\ & &
 \times \ v_{1\mu'}(v.P_1+i\hat{C})^{-1}_{v_1v_2} \ v_{2\nu'}(-v.P_3+i\hat{C})^{-1}_{v_2v_3} \ 2 C(v_3,v_5) (v.P_3+i\hat{C})^{-1}_{v_5v_4} \ v_{4\rho'} \nonumber \\ & & 
+ \ ( P_2,\nu,b \leftrightarrow  P_3,\rho,c ) \ \rbrace
\label{Ap8}
\end{eqnarray}
With the identity (\ref{Ap4}), and with $G_c(-P)=G_c(P)$, the form (\ref{Ap8}) turns into
\begin{eqnarray} 
\lefteqn{{\cal A}_{abc}^{\mu\nu\rho \ (\gamma)} = -G_r^{\mu\mu'}(P_1) \ \lbrace \ G_r^{\rho\rho'}(P_3) \ G_c^{\nu\nu'}(P_2) \  g \  i f_{abc} \ m_D^2 \ T  \nonumber}  \\ & &
 \times \ \ v_{1\mu'}(v.P_1+i\hat{C})^{-1}_{v_1v_2} \ v_{2\nu'} \ [ \ (-v.P_3+i\hat{C})^{-1}  + (v.P_3+i\hat{C})^{-1} \ ]_{v_2v_4} \ v_{4\rho'}  \nonumber 
\\ & & + \ ( \ P_2,\nu,b \leftrightarrow P_3,\rho,c \ ) \ \rbrace
\label{Ap7}
\end{eqnarray}
One recognizes ${\cal T}(1,2,3)$ defined in Eq.(\ref{b7}) in the first term of the expression (\ref{Ap7}).  The second term in (\ref{Ap7}) contains
\begin{equation}
v_{1\mu'}(v.P_1+i\hat{C})^{-1}_{v_1v_2} \ v_{2\nu'} \  (v.P_3+i\hat{C})^{-1}_{v_2v_4} \ v_{4\rho'}  = S(1,2,3)
\end{equation}
symmetric in the exchange $(P_1,\mu' \leftrightarrow P_3,\rho')$. Now, in the symmetric combination (\ref{c12}), such terms add up to zero, since $S(1,2,3) \ f_{abc}$ is cancelled by  $S(3,2,1) \ f_{cba}$  , in contrast to ${\cal T}(1,2,3)$ which obeys Eq.(\ref{b7b}).

\end{document}